\magnification=\magstep1
\font\bigbfont=cmbx10 scaled\magstep1
\font\bigifont=cmti10 scaled\magstep1
\font\bigrfont=cmr10 scaled\magstep1
\vsize = 23.5 truecm
\hsize = 15.5 truecm
\hoffset = .2truein
\baselineskip = 14 truept
\overfullrule = 0pt
\parskip = 3 truept
\def\frac#1#2{{#1\over#2}}

\nopagenumbers
%
\topinsert
\vskip 3.2 truecm
\endinsert
\centerline{\bigbfont SUPERFLUID PHASES OF TRIPLET PAIRING}
\vskip 20 truept
\centerline{\bigifont J. W. Clark}
\vskip 8 truept
\centerline{\bigrfont McDonnell Center for the Space Sciences}
\vskip 2 truept
\centerline{\bigrfont and Department of Physics, Washington University}
\vskip 2 truept
\centerline{\bigrfont St. Louis, Missouri 63130, USA}
\vskip 14 truept
\centerline{\bigifont V. A. Khodel and M. V. Zverev}
\vskip 8 truept
\centerline{\bigrfont Russian Research Center Kurchatov Institute}
\vskip 2 truept
\centerline{\bigrfont Moscow 123182, Russia}
\vskip 1.8 truecm

\centerline{\bf 1.  INTRODUCTION}
\vskip 12 truept

Triplet pairing is exemplified physically in at least two strongly
interacting quantum many-fermion systems:
\item{(i)}
In laboratories on earth:  Superfluid phases are observed when
liquid $^3$He is cooled to mK temperatures (triplet pairing of
spin--1/2 atoms) [1].
\item{(ii)}
In the quantum fluid interior of a neutron star:  It is generally
believed that the neutrons pair-condense at temperatures below
$\sim 10^9$ K (triplet pairing of spin-1/2 nucleons) [2].

There are important differences in these realizations.  In {\it dense
neutron matter},  $^1S_0$ pairing, which dominates in the inner crustal
region of the star, is quenched by the strong short-range repulsion in
the nucleon-nucleon ($NN$) interaction.  At $NN$ collision energies
corresponding to the relevant density regime, the most attractive
phase shift is found in the $^3P_2$ channel, indicating that
this state will dominate the pairing.  The $^3P_0$ phase shift
is only weakly attractive, while the $^3P_1$ phase shift is repulsive;
these channels are therefore expected to have only minor effects on
the problem.  The state dependence of the spin-orbit component of
the $NN$ interaction is responsible for this pre-eminence of the
$^3P_2$ state.  A complication is introduced by the tensor component
of the interaction, which implies a coupling of the $^3P_2$ and
$^3F_2$ channels.

On the other hand, in {\it liquid} $^3He$, the atom-atom interaction
is spherically symmetric and spin-independent to a very good
approximation.  Tensor and spin-orbit components are effectively
absent.  Thus there is no coupling to higher $L$ waves, but
the three $J$ states for $L=1$, $S=1$, namely $^3P_2$, $^3P_1$,
and $^3P_0$ now enter on an equal footing.

On balance, the neutron matter -- in spite of the complication
of tensor coupling -- is simpler than liquid $^3He$, since fewer
magnetic substates enter the problem.  Therefore we focus first
on neutron matter as a ``training ground'' for the more challenging
case of liquid $^3$He.

Quantitative, microscopic prediction of the superfluid phases and
phase diagram of systems manifesting triplet pairing (or pairing
in orbital angular momentum states beyond $S$-wave), has hitherto
eluded the best efforts of theorists.  For one thing, the popular
Ginzburg-Landau approach is restricted to temperatures near the
critical temperature $T_c$.  Secondly, iterative procedures,
routinely employed to calculate the energy gap in weakly interacting
systems with $S$-wave pairing, are afflicted with slow convergence
and uncertain accuracy when applied to the many coupled, singular nonlinear
integral equations that come into play for pairing in higher
angular momentum states.  The limitations of standard iterative
procedures become most apparent when one attempts to construct the
superfluid phase diagram of the system, which is controlled by
tiny energy splittings between different solutions of the BCS
pairing problem.

These difficulties can now be overcome by means of a recently
introduced {\sl separation method} [3--8] for solving the BCS gap
equations associated with pairing states of arbitrary angular-momentum
content.  In this paper, we implement this method to identify and
classify the spectrum of phases of the $^3P_2$--$^3F_2$ pairing model.
Nominally, the model refers to superfluid neutron matter at densities
of order twice that found in heavy nuclei.  However, the results
obtained are universal in the sense that the angular properties
of the solutions corresponding to the existing phases are independent
of the detailed nature of the system under study, and especially
its pairing interaction.  Indeed, these results can be derived
analytically and depend only on two energy scales, one
governing the overall strength of the pairing effect and the other
measuring the strength of the coupling between $P$ and $F$ states.
There exist the usual solutions involving a single value of the
magnetic quantum number $M$ (and its negative) -- obviously three
in number corresponding to $M=0,\pm 1,\pm 2$.   In addition we
find that there exist {\it ten} real multicomponent solutions.
Five of these have angle-dependent order parameters with nodes
(and therefore are of relatively high energy), while the other
five are nodeless.  In contrast to the case of superfluid
$^3$He, transitions occur between phases with nodeless order parameters.
Results for the temperature dependence of the competition between
the various phases have been obtained.  In principle, and presumably
also in practice, the same approach can be used to complete the catalog
of superfluid phases of liquid $^3$He.
\vskip 28 truept

\centerline{\bf 2.  BCS FORMALISM FOR ANY-CHANNEL PAIRING}
\vskip 12 truept

Adopting a spinor representation, the $2 \times 2$
gap matrix for the general BCS problem of pairing in a state
of triplet spin ($S=1$) and triplet isospin ($T=1$) has the expansion
$$
\Delta_{\alpha\beta}({\bf p}) =\sum_{J,L,M}\Delta_L^{JM}(p)
\left(G_{LJ}^M({\bf n})\right)_{\alpha\beta} \eqno(1)
$$
in terms of the spin-angle matrices
$$
\left(G_{LJ}^M({\bf n})\right)_{\alpha\beta}=\sum_{M_SM_L}
C^{1M_S}_{{1\over 2}{1\over 2}\alpha\beta}C^{JM}_{1LM_SM_L}
Y_{LM_L}({\bf n}) \,. \eqno(2)
$$
The coupled partial-wave gap components
$\Delta_L^{JM}(p)$ solve the set of {\sl gap equations}
$$
\eqalignno{
 \Delta_L^{JM}(p)
&=\sum_{L'L_1J_1M_1}(-1)^{1+{L-L'\over 2}}\int \int
  \langle p| V_{LL'}^J |p_1 \rangle S^{JMJ_1M_1}_{L'L_1}({\bf n}_1)
  \cr
&\qquad \times
  {\tanh \left(E({\bf p}_1) / 2T\right)\over 2 E({\bf p}_1)}
  \Delta_{L_1}^{J_1M_1}(p_1) p^2_1 dp_1d{\bf n}_1 \,.  &(3) \cr
}
$$
The { quasiparticle energy}
$E({\bf p})=
\left[\xi^2(p) + D^2({\bf p})\right]^{1/2}$
is constructed from the gap components $\Delta_L^{JM}(p)$ through
$$
D^2({\bf p}) = \sum_{LJML_1J_1M_1}
\left(\Delta_L^{JM}(p)\right)^* \Delta_{L_1}^{J_1M_1}(p)
S^{JMJ_1M_1}_{LL_1}({\bf n}) \eqno(4)
$$
together with the single-particle spectrum $\xi(p)$
of the normal Fermi liquid, often parametrized in terms of an
effective mass $M^*$.  {\sl Angular dependence} is introduced into
the quasiparticle energy $E({\bf p})$ via the spin trace
$$
S_{LL_1}^{JMJ_1M_1}({\bf n}) = {\rm Tr}
\left[\left(G_L^{JM}({\bf n})\right)^*G_{L_1}^{J_1M_1}({\bf n})\right]
\,, \eqno(5)
$$
which obviously complicates explicit solution of the system of gap equations.

The { pairing matrix elements} $\langle p| V_{LL'}^J|p_1 \rangle$ are
generated by the spin-angle expansion
$$
V({\bf  p},{\bf p}_1)
=\sum_{LL'JM}(-1)^{{L-L'\over 2}} \langle p| V_{LL'}^J|p_1 \rangle
G_{LJ}^M({\bf n})\left(G_{L'J}^M({\bf n}_1)\right)^* \eqno(6)
$$
of the totality of vertex diagrams irreducible in the particle-particle
channel.

A close inspection reveals that complete solution of this set of
equations presents awesome difficulties: note, in particular, the
coupling to the generic angular momentum labels $L_1$, $J_1$,
and $M_1$ quantum numbers via the squared-gap quantity $D^2({\bf p})$
appearing in the quasiparticle energy $E({\bf p})$.  Practical
{\it approximate} solution (which might still be extremely accurate) will
require a series of justifiable simplifications.

To wit, in dealing with neutron matter, the free-space two-body interaction
has the salient feature that the components of the central forces
nearly cancel each other, as reflected in the behavior of the
experimental $P$-scattering phases.  {\it It is assumed that this feature
is preserved by the effective interaction inside neutron matter.}
We shall later comment on the veracity of this assumption.  If it
holds, the pivotal role of the spin-orbit force in promoting the
$^3P_2$ pairing channel then implies that contributions to triplet
pairing from ``nondiagonal'' terms with $L',\, L_1\neq 1$ or $J_1\neq 2$
on the right-hand side of the set of gap equations can be evaluated
within perturbation theory, in terms of the set of {\sl principal gap
amplitudes} $\Delta_1^{2M}(p)$, with $M=0,\pm 1,\pm 2$.

In fact, if we choose to invoke {\sl time-reversal invariance}, the
problem may be treated in terms of only {\it three} complex
functions, namely $\Delta_1^{2M}(p)$ with $M=0\,,1\,,2$.

Another simplifying feature is the existence of a {\sl small
parameter} across the range of interesting pairing problems.  When
dealing with anisotropic gaps as may arise in pairing states
beyond the $S$-wave, it is helpful to define the energy gap
parameter $\Delta_F$ as the square root of the angle average
of $D^2({\bf p})$, evaluated at the Fermi surface.
{\it The ratio $d_F =\Delta_F/\epsilon_F$ of the gap to the
Fermi energy provides a small parameter for BCS pairing theory.}
For $^3P_2$ (or $^3P_2$--$^3F_2$) pairing in neutron matter, a
leading order approximation in $d_F$ is good to about one part in
$10^5$ or $10^6$.  A similar or higher accuracy is to be expected in
liquid $^3$He.
\vskip 28truept

\centerline{\bf 3. THE SEPARATION METHOD}
\vskip 12truept

The separation approach developed in Refs.\ [3-8] facilitates
essential further simplifications (linked in part to the
existence of a (very) small parameter).  In this approach,
any given pairing matrix element is expressed identically
as a separable part, plus a remainder that vanishes
when either momentum argument is on the Fermi surface.  We write
$$
\langle p | V_{LL'}^J | p_1 \rangle =
v_{LL'}^J \phi_{LL'}^J(p)\phi_{LL'}^J(p_1) + W_{LL'}^J(p,p_1)  \eqno(7)
$$
and verify that the choices
$v_{LL'}^J \equiv \langle p_F | V_{LL'}^J | p_F \rangle$ and
$\phi_{LL'}^J (p) \equiv \langle p | V_{LL'}^J | p_F \rangle / v_{LL'}^J$
produce the desired behavior when $p$ or $p_1$ hits $p_F$.  Substitution
of this identity into the set of gap equations, followed by simple
manipulation and argumentation, establishes a decomposition of the form
$$
\Delta^{JM}_L(p)=D_L^{JM}\chi_{LL'}^J (p)\qquad (|M|=0,1,2)  \eqno(8)
$$
for the general gap component and provides separate equations for
the {\sl shape factor} $\chi_{LL'}^J (p)$ (normalized to unity at
$p=p_F$) and the {\sl numerical amplitudes} $D_L^{JM}$ that
determine the angle-dependence of the gap.

The shape factor $\chi_{LL'}^J(p)$ is given by a nonsingular integral
equation involving a set of kernels proportional to the residual
interaction $W_{LL'}^J(p,p_1)$ (see Ref.\ [8] for details).
The vanishing of $W$ at the Fermi surface removes the
logarithmic singularity characteristic of BCS theory, which
is banished to the set of equations for the amplitudes
$D_L^{JM}$, which are coupled nonlinear equations for a set of
{\it numbers}.

To very high accuracy, the integral
equation for $\chi_{LL'}^J$ is {\it linear} and {\it independent}
of the amplitudes $D_L^{JM}$.  For all practical purposes
(e.g., appealing to extremely rapid convergence of small-parameter
expansions), we are free to make the replacements
$$
E({\bf p}) \rightarrow | \xi({\bf p})| \, , \qquad
\tanh (E({\bf p})/2T) \rightarrow 1  \eqno(9)
$$
in any integral involving the residual interaction $W$ as a factor.
The shape factors $\chi_{LL'}^J(p)$ are therefore determined
by the interaction $W$ and may be calculated independently of
the $D_L^{JM}$ by a matrix inversion routine.  The nonlinear
and singular aspects of the problem reside entirely in the
set of equations for the amplitudes $D_L^{JM}$, which
may also be solved numerically by standard methods (e.g. Newton-type
algorithms).  In fact, we have even been able to solve scaled versions
of the $D_L^{JM}$ equations {\it analytically} in interesting cases,
as will emerge below.

In view of these crucial simplifications revealed and expedited
by the separation transformation, it is seen that
\item{(i)}
For given $L$, $L'$, and $J$, the $\chi$ function is universal,
i.e., independent of magnetic quantum number $M$ and temperature
$T$.
\item{(ii)}
The factorization
$$
\Delta_1^{2M}(p) = D_1^{2M} \chi(p) \qquad (|M|=0,1,2) \eqno(10)
$$
may be asserted for the principal gap components, where
$\chi(p) \equiv \chi_{11}^{21}(p)$.
\item{(iii)}
Explication of the phase diagram of dense superfluid neutron
matter reduces to determination of the three amplitudes $D_1^{2M}$,
since the character of the phase diagram itself is independent
of the shape factor $\chi(p)$.
\vskip 28truept

\centerline{\bf 4. PERTURBATIVE TREATMENT}
\vskip 12truept

Now let us focus on the nondiagonal contributions to the right-hand
side of the gap equations (3), where nondiagonal means one or
more of the conditions $L'=1$, $L_1 = 1$, $J_1 =2$ is not met.
Two nondiagonal contributions are found to be of leading
importance:  The first contains the integral of the
product $V_{31}^2 S_{31}^{2M2M_1} \Delta_1^{2M_1}$
while the second contains the integral of the product
$V_{11}^2 S_{13}^{2M2M_1} \Delta_3^{2M_1}$.  There is
a single small factor in each product, namely $V_{31}$ in
the first instance and $\Delta_3^{2m_1}$.  All other
nondiagonal contributions involve two or more small factors.
Retaining the leading nondiagonal pieces and ignoring all
the others, we have what is called the $^3P_2$--$^3F_2$
{\sl pairing problem}, which is generally regarded as a
satisfactory model of superfluid neutron matter.

Rapid convergence of the pertinent nondiagonal integrals is
instrumental to success of the perturbation approach
used to treat the nondiagonal effects.  To
affirm this behavior, we observe that the dominant contributions
to these integrals come from the vicinity of the Fermi surface.
Then, if $E({\bf p})$ is significantly larger than the energy gap value
(as it will be for large $p$), $E({\bf p})$ and $| \xi({\bf p}) |$ will
practically coincide, and consequently the angular integration yields zero.

We conclude that in evaluating the nondiagonal contributions
it suffices to know the ``minor'' gap components $\Delta_3^{2M}(p)$
at $p=p_F$.  These quantities will in fact be determined
in terms of the coefficients $D_1^{2M}$.  To excellent approximation,
we may retain only the dominant contribution to the right-hand
side of the gap equations (3) that diverges like $\ln d_F$ as
$\Delta_F \rightarrow 0$.  We are then led to the desired
connection
$$
\Delta^{2M}_3(p=p_F)=\eta D^{2M}_1 \,,  \eqno(11)
$$
where $\eta=-\langle p_F| V_{13}^2|p_F \rangle/v_F$
and $v_F\equiv\langle p_F| V_{11}^2|p_F \rangle$.
Similar relations may be obtained for other minor gap components,
notably $\Delta_1^{00}$ and $\Delta_1^{1M}$.

What can we say about the parameter $\eta$, which represents
the coupling to the ``nondiagonal'' states?  If perturbation
theory is to be valid, it should be comfortably small.  For pairing
matrix corresponding directly to in-vacuum neutron-neutron interaction,
$\eta$ depends smoothly on density, varying around 0.3 in the interval
$\rho_0 < \rho < 2 \rho_0$.  Medium modification of the spin-orbit
force is probably not important.  Due to the relativistic origin
of this component, it should not be much affected by polarization or
correlation corrections.  This claim is consistent with empirical
analyses of the spin-orbit splitting in finite nuclei.

On the other hand, medium modification of the tensor force may
be more significant, especially as one approaches the density
at which pion condensation occurs.  Thus, the parameter $\eta$
is somewhat uncertain, although it is still expected to be
rather small.

To sketch out the superfluid phase diagram of the system, we need to
determine the key amplitudes $D_1^{2M}$ ($M=0,1,2$) to leading perturbative
order in the coupling $\eta$.  Exploiting the linear connection (11)
between the $D_1^{2M}$ amplitudes and the minor-component values
$\Delta_3^{2M}(p_F)$, simple manipulations applied to the coupled
gap equations (1) at $p=p_F$ yield
$$
D^{2M}_1+v_F\sum_{M_1}D^{2M}_1\int\int\phi(p)
{\tanh {\left( E_0({\bf p})/ 2T \right)}\over
2E_0({\bf p})} S^{2M2M_1}_{11}({\bf n})\chi(p)p^2 dp
d{\bf n}= \eta v_F r_M  \eqno(12)
$$
with $\phi(p)\equiv\langle p| V^2_{11}|p_F \rangle/v_F$,
$E_0({\bf p}) \equiv E({\bf p}; \eta=0)$, and
$$
r_M =\sum_{M_1}D_1^{2M_1}\int\int
\left[S^{2M2M_1}_{31}({\bf n})+S^{2M2M_1}_{13}({\bf n})\right]
{\tanh {\left(E_0({\bf p})/2T\right)}\over 2E_0({\bf p})}
  p^2 dp d{\bf n} \,. \eqno(13)
$$
We restrict the search for solutions of these equations to those
with real coefficients $D_1^{2M}$, reasoning that solutions with
complex $D$ amplitudes will lie at energies high enough to make
them physically irrelevant.

Inserting the explicit form of $S^{2M2M_1}_{11}({\bf n})$ from
Eq.~(5) into Eq.~(12), we derive a system of three equations for the
gap value $\Delta_F$ and the two ratios
$$
\lambda_1 = D_1^{21}/D_1^{20} \sqrt{6} \quad {\rm and} \quad
\lambda_2 = D_1^{22}/D_1^{20} \sqrt{6} \,,  \eqno(14)
$$
which serve generally to determine the angular dependence of the
gap (or alternatively its composition with respect to the magnetic
quantum number $M$).  In a notation and form compatible with the
earlier treatment of the pure $^3P_2$ problem [4,6], these
{\sl basic equations} read
$$
\eqalignno{
  \lambda_2+ v_F\left[\lambda_2(J_0+J_5) -\lambda_1 J_1 -J_3\right]&=
\eta v_F r_2\,, &(15a)\cr
 \lambda_1+ v_F\left[-(\lambda_2+1)J_1+\lambda_1(J_0+4J_5+2J_3)/4\right]&=
\eta v_F r_1\,, &(15b)\cr
1+ v_F\left[-(\lambda_2 J_3+\lambda_1 J_1)/3 +J_5\right]&=\eta v_F r_0 \,,
&(15c)\cr }
$$
with
$$
J_i = \int\int f_i(\theta,\varphi)
\phi(p){\tanh\left(E_0({\bf p})/ 2T\right)\over 2E_0({\bf p})}\chi(p)
{p^2dpd{\bf n}\over 4\pi}\,, \eqno(16)
$$
$$
f_0=1-3z^2\,, \quad f_1=3xz/2\,, \quad f_3= 3(2x^2+z^2-1)/2 \,, \quad
f_5=(1+3z^2)/2 \,, \eqno(17)
$$
$$
z=\cos \theta \,, \quad x=\sin\theta\cos\varphi\,,
\quad y=\sin\theta\sin\varphi \,. \eqno(18)
$$
Upon setting $\eta = 0$, Eqs.~(15a)--(15c) collapse to Eqs.~(12)--(14)
of Ref.~[4], which refer to the case $\kappa_1 = \kappa_2 = 0$
and were solved analytically in Refs.~[4,6].

If we had {\it not} used the separation method, we would have been
faced at this point with a system of coupled singular nonlinear integral
equations for a set of {\it functions}, not {\it numbers}.  The great
advantage gained is transparent.
\vskip 28truept

\centerline{\bf 5. SEARCHING FOR MULTICOMPONENT SOLUTIONS}
\vskip 12truept

The basic equations (15a)--(15c) possess the familiar {\sl one-component}
solutions corresponding to definite $|M|$, i.e., $|M| = 0, 1, 2$.
In addition, there exist {\sl multicomponent} solutions whose
structure and spectrum we seek to establish, via a {\it two-step
process}.  We first note that $J_5$, which is the only $J_i$ integral
containing a principal term going like $\ln(\epsilon_F/\Delta_F)$,
is responsible for the gap magnitude $\Delta_F$.  On the other hand,
$J_5$ is irrelevant to the phase structure.

\noindent
{\sl Step 1.}  Thus, we begin by eliminating $J_5$ from
the first pair (15a)--(15b) of basic equations and reduce
the number of $J_i$ integrals in each of the pair to two:
$$
\eqalignno{
(\lambda_2+1)[  3\lambda_1(\lambda_2+1)J_0 -
           2(\lambda^2_1-2\lambda_2^2+6) J_1] &=\eta B_1
   \,, &(19a) \cr
(\lambda_2+1)[ (\lambda_1^2-4\lambda_2)J_1 +
                    \lambda_1(\lambda_2+1)J_3] &=\eta B_2 \,, &(19b) \cr
}
$$
$$
\eqalignno{
B_1&=2\lambda_1(2\lambda_2+3)r_2-4(\lambda^2_2-3)r_1-
6\lambda_1(\lambda_2+2)r_0 \,, \cr
B_2&=-\lambda_1 r_2+4\lambda_2r_1-3\lambda_1\lambda_2 r_0 \,.  &(20)\cr
}
$$

\noindent
{\sl Step 2.}  Assuming $\lambda_2 \neq 1$, we perform the
rotation
$$
(x,z)=(-t\sin\vartheta+u\cos\vartheta,\, t\cos\vartheta+u\sin\vartheta)
\eqno(21)
$$
and choose the angle $\vartheta$ so as to remove the integral $J_1$
from the ``sanitized'' first pair of basic equations, (19a)--(19b).
Specifically, if $\zeta=\tan\vartheta$ is chosen to obey the algebraic
equation
$$
\lambda_1\zeta^2-(\lambda_2-3)\zeta-\lambda_1=0 \,, \eqno(22)
$$
then substitution of the transformed integrals $J_i$ converts the key
pair of equations (19a)--(19b) into
$$
\eqalignno{
(\lambda_2+1)[A_1J_0+A_2J_3]&=\eta B_1 \,, &(23a) \cr
(\lambda_2+1)[A_1J_0+A_2J_3]&=-2 \eta B_2 \,, &(23b)  \cr
}
$$
with
$$
\eqalignno{
A_1&= {3\over
2}\lambda_1(1+\lambda_2)(2-\zeta^2)
-{3\over 2}(\lambda_1^2-2\lambda_2^2+6)\zeta \,, \cr
A_2&=-3\lambda_1(1+\lambda_2)\zeta^2-(\lambda_1^2 -2\lambda_2^2+6)\zeta
\,.  &(24)\cr }
$$

We observe that the left-hand members of Eqs.~(23a) and (23b) are
{\it identical!}  It is in fact just this coincidence that leads to
the striking {\sl universalities} of the pure $^3P_2$ pairing problem
first revealed in Ref.~[4].  Of course, in this problem, which
corresponds to $\eta=0$, we have the special circumstance that
the right-hand side of each equation vanishes identically, so
that Eqs.~(23a) and (23b) coincide and yield only the single
constraint
$$
(\lambda_2+1)[A_1J_0+A_2J_3]=0  \eqno(25)
$$
on the parameters $\lambda_1$ and $\lambda_2$.
\topinsert
\input psfig.sty
\centerline{\hskip-10mm\psfig{figure=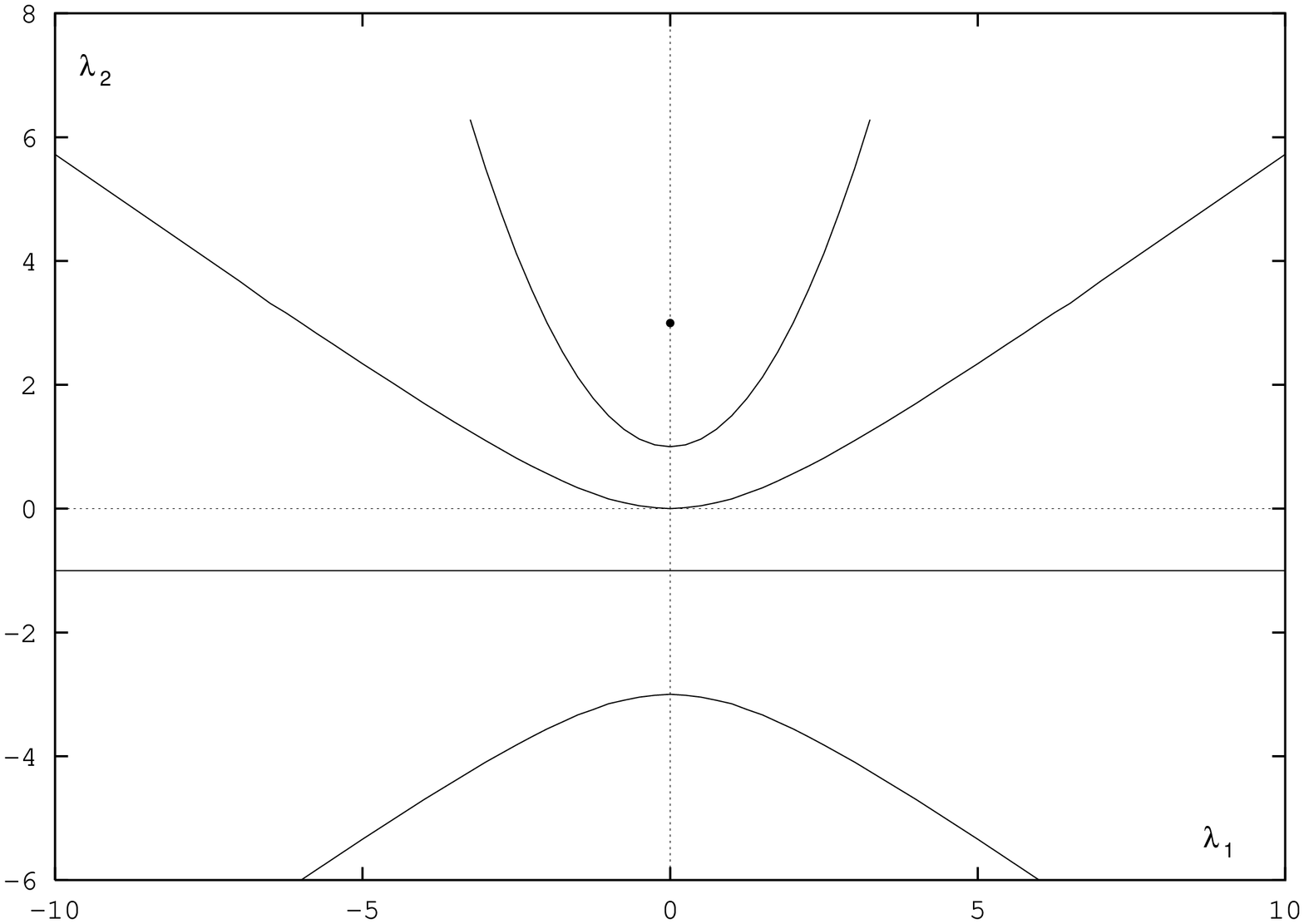,height=11truecm,width=15truecm,angle=270}}
\vskip .3truecm
\noindent
{\bf Figure 1.}
Solutions for the coefficient ratios $\lambda_1$ and $\lambda_2$
in the case $\eta \equiv 0$, corresponding to the uncoupled $^3P_2$
pairing problem solved in Refs.~[4,6].  The particular (point) solution
($\lambda_1=0$, $\lambda_2=3$) is indicated by the solid dot and
the degenerate solutions by the solid curves.
\vskip 12truept
\endinsert

It ensues that the solutions of the pure $^3P_2$ pairing
problem display remarkable universalities as expressed in
two kinds of degeneracies, independently of temperature,
density, and details of the in-medium interaction:
\item{$\bullet$}
{\sl Energetic degeneracy}.  There exists an {\it upper} group of
states, degenerate in energy, whose angle-dependent order
parameters {\it have nodes}, and a {\it lower} group with {\it nodeless}
order parameters.  Relative to the absolute pairing energy,
the splitting between upper and lower states is small
(of order 2\% in the neutron-matter problem).
\topinsert
\vskip -4.3truecm
\input psfig.sty
\centerline{\hskip5mm\psfig{figure=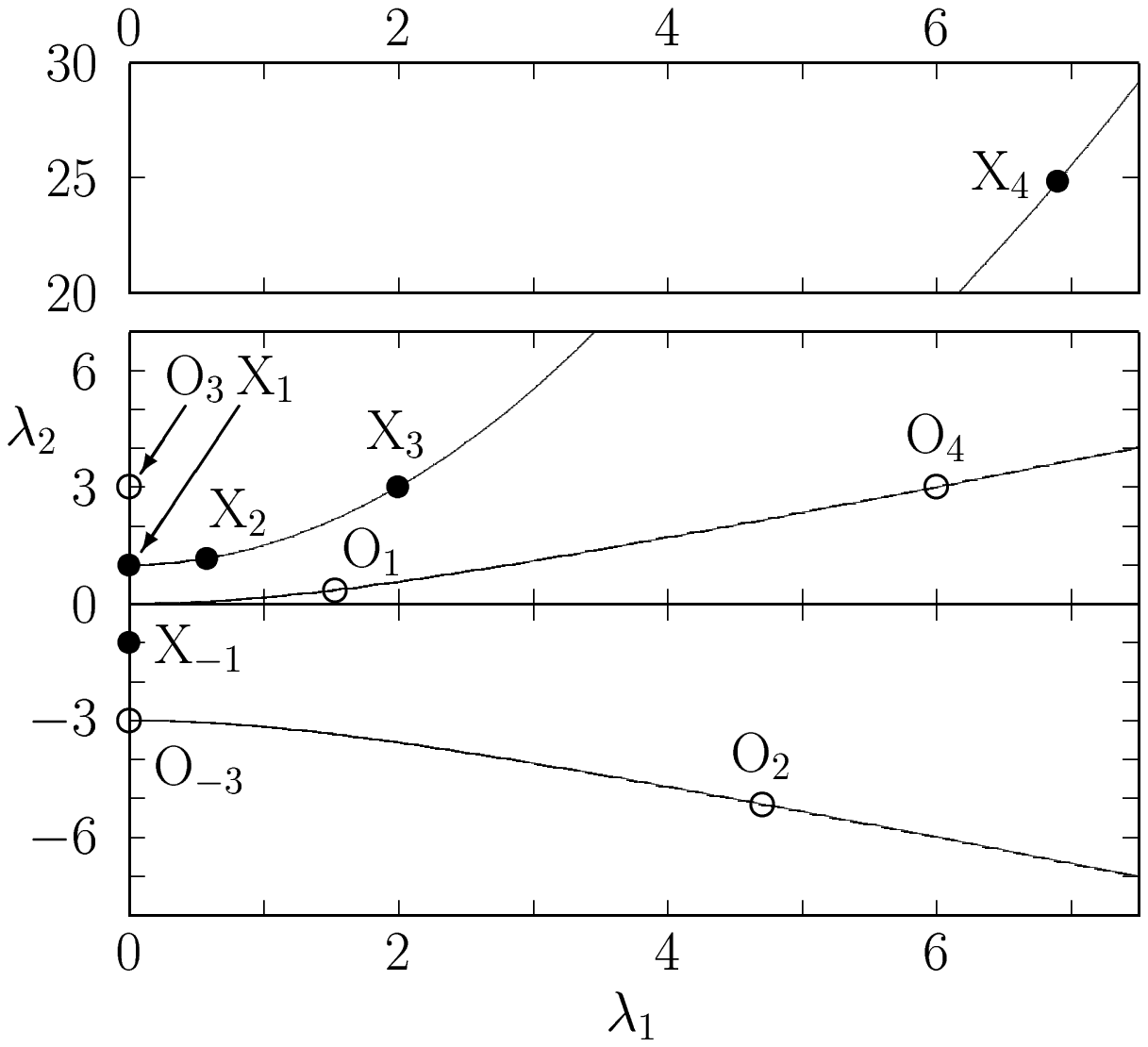,height=28truecm,width=20truecm,angle=0}}
\vskip -11.4truecm
\noindent
{\bf Figure 2.}
Multicomponent solutions of the $^3P_2$ pairing problem, defined
by the coefficient ratios $\lambda_1$ and $\lambda_2$.  Solutions
with nodeless (respectively, node-bearing) order parameters
in the case of pure $^3P_2$ pairing are indicated by open
(respectively, filled) circles.
\vskip 12truept
\endinsert
\item{$\bullet$}
{\sl Parametric degeneracy}.  The multicomponent solutions,
which satisfy the spectral condition [4,6]
$$
(\lambda^2_1+2-2\lambda_2)(\lambda^2_1-2\lambda^2_2-6\lambda_2)=0 \,,\eqno(26)
$$
display a parametric degeneracy with respect to the parameters
$\lambda_1$ and $\lambda_2$.  Thus, as seen in Fig.~1 (which was
constructed analytically), the multicomponent solutions generally
define {\it curves} rather than {\it points} in the
$(\lambda_1,\lambda_2)$ plane.
\vskip 28truept

\centerline{\bf 6. BREAKING THE PARAMETRIC DEGENERACY}
\vskip 12truept

The strong parametric degeneracy intrinsic to the uncoupled $^3P_2$
pairing problem is lifted in the case of $^3P_2$--$^3F_2$ pairing,
i.e., at $\eta\neq 0$.  With the coupling turned on,
the true solutions of the problem are represented by a set of
isolated {\it points} in the $(\lambda_1,\lambda_2)$ plane.

From the analytical standpoint, the underlying ``mechanism'' runs as follows.
Upon equating the right-hand members of the double-sanitized pair
of basic equations (23a)--(23b) in the small--$|\eta|$
limit ($|\eta|$ infinitesimal), one obtains an additional relation
between the parameters $\lambda_1(\eta=0)$ and $\lambda_2(\eta=0)$:
$$
\lambda_1 r_2-(\lambda_2-3)r_1-3\lambda_1 r_0=0 \,.\eqno(27)
$$
This relation supplements the spectral condition (26)
nontrivially and {\it removes the parametric degeneracy}.
The system formed by (27) and (26) is solved by applying
the same rotation in $x-z$ coordinates as introduced previously.
After some algebra, one may then obtain the full set of solutions
of the coupled-channel $^3P_2$--$^3F_2$ pairing problem.
A salient feature of these solutions, made explicit via the
separation scheme, is their virtually complete {\it independence}
of the temperature $T$.

We summarize the results of this analysis with a catalog of the
possible solutions for $^3P_2$--$^3F_2$ pairing described in the
BCS framework.  One simplification, already evident in Fig.~1
but also quite general, is that the relevant pairing energies
are independent of the sign of $\lambda_1$, so that we only need
to consider $\lambda_1 > 0$.  The modified picture in the
$(\lambda_1,\lambda_2)$ plane is displayed in Fig.~2.

First of all, there remain the {\it three} well-known {\sl single-component
solutions} with $|M| = 0$, 1, or 2.  Beyond these, the collection of
{\sl unitary solutions} of the $^3P_2$--$^3F_2$ pairing problem contains
{\it ten} {\sl multicomponent solutions}, corresponding to more
complicated superfluid phases.

{\it Five} of these additional solutions, denoted $O_k$ ($k=1,\ldots,5$),
have {\it nodeless} order parameters and include:
\item{(a)}
{\it Two} {\sl two-component solutions} $O_{\pm3}$, identical to those
found in the pure $^3P_2$ pairing problem, with $\lambda_1=0$ and
$\lambda_2=\pm 3$.
\item{(b)}
{\it Three} {\sl three-component solutions}:
\itemitem{$\circ$}
Two of them, $O_1$ and $O_4$, are associated with the upper branch
of $\lambda^2_1-2\lambda^2_2-6\lambda_2=0$ and have
$\lambda_2=3(\sqrt{21}{-}4)/5$ and $\lambda_2=3$, respectively.
\itemitem{$\circ$}
The third, $O_2$, is associated with the lower branch of the
same equation and has $\lambda_2=-3(\sqrt{21}+4)/5$.

The other five solutions, denoted by $X_k$ ($k=1, \ldots,5$)
do possess {\it nodes}.
\item{(a)}
{\it Two} {\sl two-component solutions} $X_{\pm1}$, again identical to
those found in the pure $^3P_2$ pairing problem, with
$\lambda_1=0$ and $\lambda_2=\pm 1$.
\item{(b)}
{\it Three} {\sl three-component solutions}, $X_2$, $X_3$, and $X_4$,
associated with the parabola $\lambda_2=\lambda^2_1/2+1$ and
having $\lambda_2=13-2\sqrt{35}$, $\lambda_2=3$, and
$\lambda_2=13+2\sqrt{35}$, respectively.

These general features of the spectrum of solutions of the
$^3P_2$--$^3F_2$ problem are expected to persist even if $|\eta|$
is not so small.
\vskip 28truept

\centerline{\bf 7. THE BATTLE BETWEEN PHASES}
\vskip 12truept

To complete the phase diagram of superfluid neutron matter,
we need the gap values $\Delta_F$ for the various phases.
These are found through the third basic equation (15c),
which involves $J_5$.  To determine which phase wins the
competition at a given temperature $T$, we compare the free-energy
shifts
$$
F_s=-\int_0^g \Delta^2_F(g'){dg' \over (g')^2} \eqno(25)
$$
due to pairing in the corresponding superfluid states,
where $g$ is the relevant coupling constant.

At low $T$, the only viable contestants are the solutions with
nodeless order parameters, since the other solutions lie too
high in energy.  The separation between the two groups of
states simply cannot be bridged if the value of $|\eta|$ remains
rather small.

As documented in the last section, the parametric degeneracy
of the $^3P_2$ pairing problem in the $\lambda_1-\lambda_2$ plane,
embodied in the spectral relation (26), is completely eradicated
when the $\eta$-coupling is switched on.  On the other hand, the
{\it energetic} degeneracy between the different superfluid phases
is only partially lifted.  Specifically, the spectrum of pairing
energies decays into several groups of nearly degenerate states:
The $O_1$ and $O_2$ phases form the lowest-energy group, followed by
the phases $O_{\pm 3}$ along with the one-component phase with $M=0$,
and so on.
\topinsert
\vskip -4.5truecm
\input psfig.sty
\centerline{\hskip5mm\psfig{figure=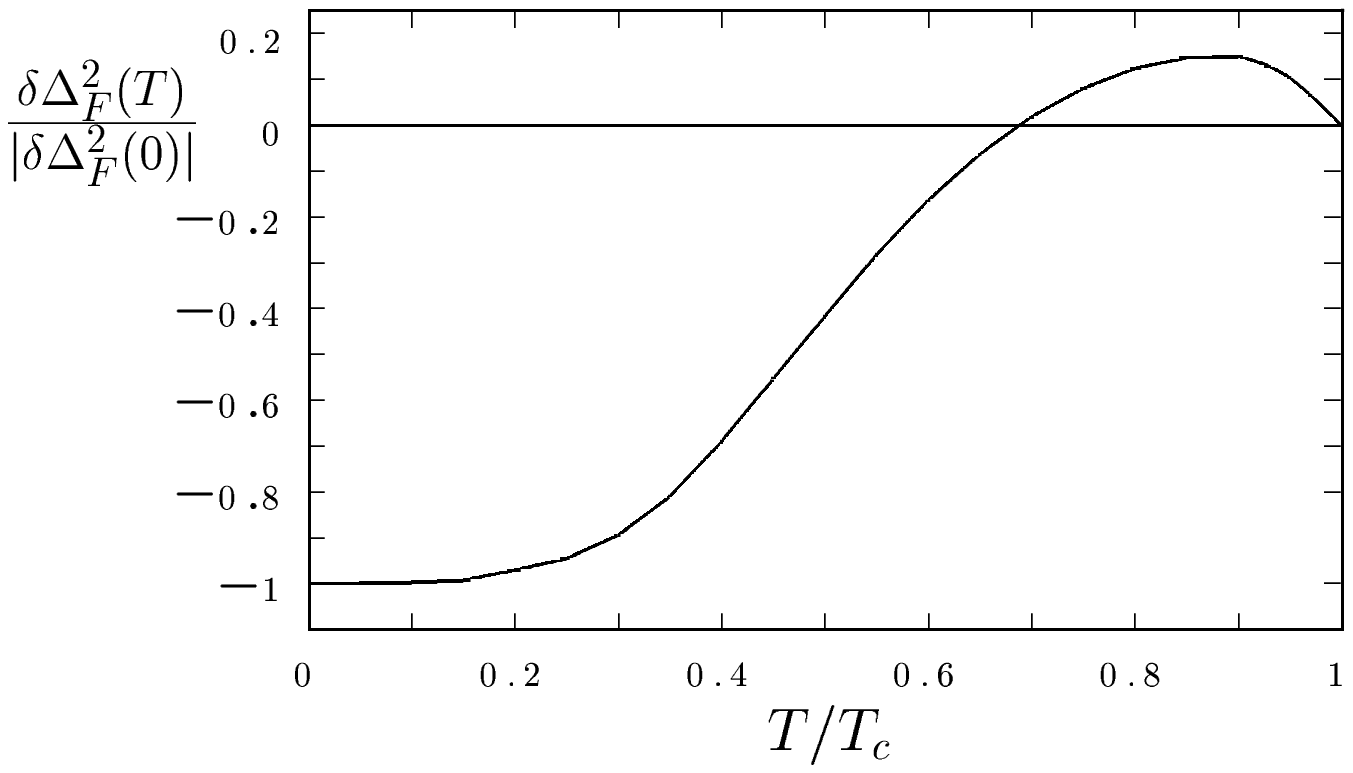,height=28truecm,width=20truecm,angle=0}}
\vskip -13.9truecm
\noindent
{\bf Figure 3.}
Temperature dependence of the splitting between the phase
$O_1$ (or $O_2$) and the phases $O_{\pm 3}$ (or the one-component
phase with $M=0$), in terms of the difference $\delta \Delta_F^2(T)$
between the respective $\Delta_F^2(T)$ values, measured relative
to $|\delta \Delta_F^2(T=0)|$.  To set the energy scales involved,
the latter quantity is around 2\% of $\Delta_F^2$ for $\eta=0.3$,
while the gap parameter $\Delta_F(T=0)$ itself reaches a maximum value
near 0.44 MeV at $k_F \simeq 2.1 $ fm$^{-1}$ when the pairing interaction
is given by the Argonne $v_{18}$ potential and free normal-state
single-particle energies are employed [8].
\vskip 12truept
\endinsert

Raising the temperature from $T=0$ toward the critical temperature
$T_c$, the splitting between the two groups lowest in energy shrinks,
until, at $T\simeq 0.7\,T_c$, their roles are interchanged and transitions
occur.  This behavior is quantified in Fig.~3.  We observe that
the analogy with the A-B phase transition in superfluid $^3$He is
imperfect.  Whereas the order parameter of the A-phase solution
has nodes while that of the B-phase does not, the transition in
neutron matter takes place between phases with nodeless order
parameters.
\vskip 28truept

\centerline{\bf 8. CONCLUSIONS}
\vskip 12truept

We close with some details and caveats and examine some prospects
for observable implications of our findings.

First, there is the matter of unresolvable fine structure in the
phase diagram.  Introduction of the $^3P_0$ and $^3P_2$ pairing
channels would entail a further lifting of energy degeneracies,
but current ignorance of the in-medium interaction
precludes reliable estimation of these effects.

Secondly, we point out that the region near the critical temperature
$T_c$ may admit subtle phase transitions not uncovered by our analysis.
In this regime, strong-coupling corrections [9,10] can no longer be
neglected in deciding the contest between phases.  Also, the
character of the phase diagram may be influenced significantly by
external magnetic fields, a fact not considered in our treatment.
And further, the phenomenon of fermion condensation [11], occurring
as a precursor to pion condensation [12-14], may provide an exotic
source of phase transitions in superfluid neuron matter.

In summary, it has been established that the superfluid phase
diagram of dense neutron matter can in principle exhibit several
triplet superfluid phases (at least 13!).  Transitions between the
different phases are expected to occur as a young neutron star cools.
Since the gap value changes in these transitions, their
occurrence may ultimately be detected in the thermal history
and/or the rotational dynamics of the star.  Such transitions may
produce a significant variation of the moment of inertia of the star,
or alterations of the distribution of the angular momentum between
the crust and the vortex system, resulting in a change of the
star's angular velocity.

The existing analysis of triplet pairing in neutron matter provides a
firm foothold for the more ambitious project of a complete characterization
of the repertoire of superfluid phases of liquid $^3$He, which is
even more complex, promising a richness of phenomena not yet
revealed by experiment or theory.  Since $^3P_2$, $^3P_1$, and $^3P_0$
states all enter the picture, one will be faced with {\it nine} nonlinear
equations for the relevant gap components (in contrast to the {\it five}
encountered in the pure $^3P_2$ case).  Nevertheless, the separation
method will continue to provide an incisive tool of analysis.

Thus, the procedure applied here may in principle be extended
to count all possible solutions of the set of nine BCS equations
for the gap function of superfluid $^3$He.  We may begin by severing
the couplings between $^3P_2$, $^3P_1$, and $^3P_0$ components, which
facilitates a reference analytic solution.  With this solution at hand,
we may restore the channel couplings and solve the resulting set of
equations numerically step by step, under increase of the coupling
parameters.  Attention then turns to (i) verification of the structure
of the A-phase predicted by ABM [15,16], (ii) evaluation of
the difference between free energies of A and B phases vs.~temperature
and pressure, and the barrier between the two phases (important for
understanding the huge delay of the phase transition between the
supercooled A-phase and the B-phase [17--19]),
(iii) investigation of strong-coupling corrections beyond the Ginzburg-Landau
approximation, and (iv) examination of the phase diagram of liquid $^3$He
in an external magnetic field.
\vskip 26truept

\centerline{\bf ACKNOWLEDGMENTS}
\vskip 10 truept

The research described herein was supported by the U.\ S.\ National Science
Foundation under Grant No.~PHY-9900713 (JWC and VAK), by the McDonnell
Center for the Space Sciences (VAK), and by Grant No.~00-15-96590 from
the Russian Foundation for Basic Research (VAK and MVZ).  JWC is grateful
to the US Army Research Office--Research Triangle Park for travel
support through a grant to Southern Illinois University--Carbondale.
We thank A.~Sedrakian, I.~I.~Strakovsky, G.~E.~Volovik, and
D.~N.~Voskresensky for stimulating
and informative discussions.
\vskip 26 truept

\centerline{\bf REFERENCES}
\vskip 10 truept
\item{[1]}
D.~D.~Osheroff, {\it Rev.~Mod.~Phys.}~{\bf 69}, 667 (1997).
\item{[2]}
T.~Takatsuka and R.~Tamagaki,
{\it Prog.~Theor.~Phys.~Suppl.}~{\bf 112}, 27 (1993).
\item{[3]}
V.~A.~Khodel, V.~V.~Khodel, and J.~W.~Clark, {\it Nucl.~Phys.}~{\bf A598},
390 (1996).
\item{[4]}
V.~A.~Khodel, V.~V.~Khodel, and J.~W.~Clark, {\it Phys.\ Rev.\ Lett.} {\bf 81},
3828 (1998).
\item{[5]}
J. W. Clark, V. A. Khodel, and V. V. Khodel, in {\it Condensed Matter
Theories}, Vol.~13, ed.\ J. da Providencia and F. B. Malik (Nova Science
Publishers, Commack, NY, 1998).
\item{[6]}
T. Lindenau, M. L. Ristig, and J. W. Clark, in {\it Condensed Matter
Theories}, Vol.~14, ed.\ D. J. Ernst, I. Perakis, and S. Umar
(Nova Science Publishers, Commack, NY, 1999), pp.\ 131-139.
\item{[7]}
J. W. Clark, V. V. Khodel, and V. A. Khodel, in {\it Condensed Matter
Theories}, Vol.~15, ed.\ G. S. Anagnostatos, R. F. Bishop, K. A. Gernoth,
J. Ginis, and A. Theophilou (Nova Science Publishers, Huntington, NY,
2000), pp.\ 1-12.
\item{[8]}
V.~V.~Khodel, V.~A.~Khodel, and J.~W.~Clark, {\it Nucl.\ Phys.}
{\bf A679}, 827 (2001).
\item{[9]}
D.~Vollhardt and P.~W\"olfle, {\it The Superfluid Phases of Helium 3}
(Taylor \& Francis, London, 1990).
\item{[10]}
J.~A.~Sauls and J.~W.~Serene, {\it Phys.\ Rev.\ D} {\bf 17}, 1524
(1978).
\item{[11]}
V. A. Khodel and V. R. Shaginyan, {\it JETP Lett.} {\bf 51}, 553
(1990).
\item{[12]}
D.~N.~Voskresensky, V.~A.~Khodel, M.~V.~Zverev, and J.~W. Clark,
{\it Ap.~J.} {\bf 533}, L127 (2000).
\item{[13]}
J. W. Clark, V. A. Khodel, and M. V. Zverev (2001) {\it Physics of
Atomic Nuclei (Yad.\ Fiz.)} {\bf 64}, 619-626 (2001).
\item{[14]}
J. W. Clark, V. A. Khodel, and M. V. Zverev, in {\it Condensed Matter
Theories}, Vol.~16, ed.\ S.~Hernandez and J. W. Clark (Nova Science
Publishers, Huntington, NY, 2001), pp. 263-273.
\item{[15]}
P. W. Anderson and P. Morel, {\it Phys.\ Rev.} {\bf 123}, 1911 (1961).
\item{[16]}
P. W. Anderson and W. F. Brinkman, {\it Phys.~Rev.~Lett.} {\bf 30},
1108 (1973).
\item{[17]}
A. J. Leggett, {\it J. Low Temp. Phys.} {\bf 87}, 571 (1992).
\item{[18]}
D. D. Osheroff and M. C. Cross, {\it Phys. Rev. Lett.} {\bf 38}, 905 (1977).
\item{[19]}
D. S. Buchanan, G. W. Swift, and J. C. Wheatley, {\it Phys.\ Rev.\ Lett.}
{\bf 57}, 341 (1986).
\end